\documentclass[11pt,a4paper]{article}
\usepackage{jcappub}
\usepackage{xspace}

\usepackage{color}

\usepackage[normalem]{ulem}

\usepackage{algorithm}
\usepackage[noend]{algpseudocode}

\makeatletter
\def\BState{\State\hskip-\ALG@thistlm}
\makeatother

\PassOptionsToPackage{usenames,dvipsnames,svgnames}{xcolor}
\usepackage{subfigure}
\usepackage{wrapfig}
\usepackage{tikz}
\usetikzlibrary{shapes,snakes,arrows,positioning,automata, matrix, fit, positioning}

\usepackage{booktabs}
\usepackage{amstext}
\usepackage{amssymb}
\usepackage{graphicx}
\usepackage{latexsym}
\usepackage{mathrsfs}
\usepackage{calrsfs}
\usepackage{amsfonts}

{\count255=\time\divide\count255 by 60 \xdef\hourmin{\number\count255}
  \multiply\count255 by-60\advance\count255 by\time
  \xdef\hourmin{\hourmin:\ifnum\count255<10 0\fi\the\count255}}

\def\vev#1{ \left\langle #1 \right \rangle }

\def\emul{_\mathrm{em}}
\def\f{f\emul}
\def\pem{p\emul}
\def\df{\Delta f}
\def\Xtr{X_\mathrm{train}}
\def\Xcv{X_\mathrm{CV}}
\def\Xsamp{X_\mathrm{samp}}
\def\Ycv{Y_\mathrm{CV}}
\def\Ytr{Y_\mathrm{train}}
\def\x{\mathbf{x}}
\def\y{\mathbf{y}}
\def\Ntr{N_\mathrm{tr}}
\def\Nsamp{N_\mathrm{samp}}
\def\L{\mathcal L}

\def\lem{\ell\emul}
\def\Dp{\Delta p}
\def\DL{\Delta \L}
\def\Dl{\Delta \ell}
\def\DZ{\Delta Z}
\def\Zem{Z\emul}
\def\nuis{\boldsymbol{\theta}_N}

\def\|{\, | \,}

\def\func#1{{\it #1}}
\def\file#1{{\tt #1}}

\newcommand{\kd}{$k$--d\xspace}
\providecommand{\e}[1]{\ensuremath{\times 10^{#1}}}

\begin{document}

\title{Learn-As-You-Go Acceleration of Cosmological Parameter Estimates}

\author[]{Grigor Aslanyan,}
\author[]{Richard Easther,}
\author[]{and Layne C. Price}

\affiliation[]{Department of Physics, University of Auckland, Private Bag 92019, Auckland, New Zealand}

\emailAdd{g.aslanyan@auckland.ac.nz}
\emailAdd{r.easther@auckland.ac.nz}
\emailAdd{lpri691@aucklanduni.ac.nz}

\date{\today}

\abstract{
Cosmological analyses can be accelerated by approximating slow calculations using a training set, which is either precomputed or generated dynamically. However, this approach is only safe if the approximations are well understood and controlled. This paper surveys issues associated with the use of machine-learning based emulation strategies for accelerating cosmological parameter estimation. We describe a learn-as-you-go algorithm that is implemented in the \textsc{Cosmo++} code\footnote{\url{http://cosmopp.com}} and (1) trains the emulator while simultaneously estimating posterior probabilities; (2)  identifies unreliable estimates, computing the exact numerical likelihoods if necessary; and (3) progressively learns and updates the error model as the calculation progresses.  We explicitly describe and model the emulation error and show how this can be  propagated into the posterior probabilities.  We apply these techniques to the \emph{Planck} likelihood and the calculation of $\Lambda$CDM posterior probabilities. The computation is significantly accelerated without a pre-defined training set and uncertainties in the posterior probabilities are subdominant to statistical fluctuations. We have obtained a speedup factor of $6.5$ for Metropolis-Hastings and $3.5$ for nested sampling. Finally, we discuss the general requirements for a credible error model and show how to update them on-the-fly.}
\maketitle

\section{Introduction}\label{intro_sec}

Constraining cosmological models using  experimental data typically requires many evaluations of the data likelihood, which can be very slow.
For instance, the analysis of galaxy surveys typically requires an estimate of the non-linear matter power spectrum and using $N$-body simulations to achieve the desired accuracy would be computationally prohibitive. Even calculating data likelihoods for cosmic microwave background (CMB) data, such as WMAP~\cite{Hinshaw:2012aka} and \emph{Planck}~\cite{Ade:2013kta}, requires the angular power spectra, which must be obtained through the evolution of Boltzmann codes like \textsc{Camb}~\cite{Lewis:1999bs} and \textsc{Class}~\cite{Lesgourgues:2011re,Blas:2011rf} that take significant fractions of a second to calculate for each set of parameters in a long Markov chain.

A common approach to speeding up computationally expensive calculations is to employ a statistical model that  \emph{emulates} an expensive calculation using  a \emph{training set} from which  results are empirically approximated.\footnote{Following the statistics literature, \emph{e.g.}, Ref.~\cite{O’Hagan20061290}, we use the term \emph{emulator} to refer to a statistical or deterministic model that approximates the output of a \emph{simulation} code in a computationally efficient fashion.} Several such algorithms for cosmological data analysis have been described, and associated codes released~\cite{Kaplinghat:2002mh,Jimenez:2004ct,Ringeval:2013lea,Petri:2015ura}. Recombination and the CMB likelihood calculations are emulated by \textsc{Rico}~\cite{Fendt:2008uu} and \textsc{Pico}~\cite{Fendt:2006uh,Fendt:2007uu} respectively, while the likelihoods themselves are cached and re-used for interpolation with \textsc{InterpMC} \cite{Bouland:2010mv}.
The \textsc{Coyote Universe} suite \cite{Heitmann:2008eq,Heitmann:2009cu,Lawrence:2009uk,Schneider:2010gv,Kwan:2012nd,Kwan:2013jva,Heitmann:2013bra} emulates the matter power spectrum using Gaussian processes and $N$-body simulations, with significant attention placed on the optimal construction of training sets.

Although emulators can significantly accelerate calculations, it can be difficult to estimate and control the errors they induce.  Consequently, emulators can only be used if the  methodology is known to be highly reliable. Two obvious but unattractive solutions to this dilemma are to make direct comparisons with  the final exact calculation or to build very dense training sets. Further, because the specific regions of parameter space to be explored may not be known in advance, the training must either span a large region of parameter space or be extended dynamically in order to ensure accuracy.  Obviously, the  advantages of emulation are eroded if significant computational resources are needed to train the emulator or the uncertainty associated with the emulation is not well-controlled.

An approach to solving this problem was demonstrated in Refs~\cite{Auld:2006pm,Bouland:2010mv,Graff:2011gv,Graff:2013cla}, where an external statistical sampling method was coupled to a machine learning algorithm to dynamically refine a statistical emulator. This results in a training set that is dense only in areas of parameter space where it is needed, while simultaneously evaluating a model's posterior. We call methods of this sort \emph{learn-as-you-go emulators}, since they both accelerate a  calculation that would be performed in any case and build a training set for future use.

In this paper we introduce a flexible and powerful learn-as-you-go algorithm that incorporates an adaptive and trainable error model for the differences between  emulated and ``exact''  cosmological likelihood evaluations.\footnote{A fully documented implementation of the algorithm is included in  \textsc{Cosmo++}~\cite{Aslanyan:2013opa},  available at \href{http://cosmopp.com}{http://cosmopp.com}. The code is modular and could  easily be applied to similar problems in other fields. In addition, we  provide a fast version of the \emph{Planck} likelihood module that uses this algorithm (see Sect.~\ref{cmb_sec} and Appendix~\ref{app_impl} for further details).}  We use well-tested sampling methods to  simultaneously evaluate model posteriors, generate the training set, and  update the error model by cross-validation.  Critically, the local emulation errors are propagated through the calculation, allowing us to estimate the error in the posterior probability $p(\x \| D)$ for data $D$ and parameter $\x$ induced by our use of emulation.  Consequently, the uncertainty in the posteriors is modelled explicitly, quantifying the robustness of the estimated posterior.

\begin{figure}
\centering
\begin{tikzpicture}[
  myscope/.style={node distance=1em and 0em},
  mymatrix/.style={matrix of nodes, nodes=block,
    column sep=2em,
    row sep=1em},
  block/.style={draw=blue, thick, fill=blue!10, rounded corners,
    minimum width=6em,
    minimum height=2em},
  vhilit/.style={draw=red,
    inner sep=0.85em},
  hhilit/.style={draw=black, thick, densely dotted,
    inner xsep=2em,
    inner ysep=.5em},
  line/.style={thick, -latex, shorten >= 2pt}
  ]

  \matrix[mymatrix] (mx) {
    & Parameter space sampler \\
    & & \\
    & Learn-as-you-go \\
    & & \\
    Error model  & Emulator & Likelihood \\
  };

  \draw[line] (mx-1-2) -- (mx-3-2);

  \draw[line] (mx-3-2) -- (mx-5-1);
  \draw[line] (mx-3-2) -- (mx-5-2);
  \draw[line] (mx-3-2) -- (mx-5-3);

  \draw[line] (mx-5-1) -- (mx-5-2);

\end{tikzpicture}
\caption{\label{diagram_fig} The relationship between different classes in the object-oriented implementation of our learn-as-you-go algorithm. An arrow pointing from one block to another indicates the first class ``knows about'' or uses the output of the second.}
\end{figure}
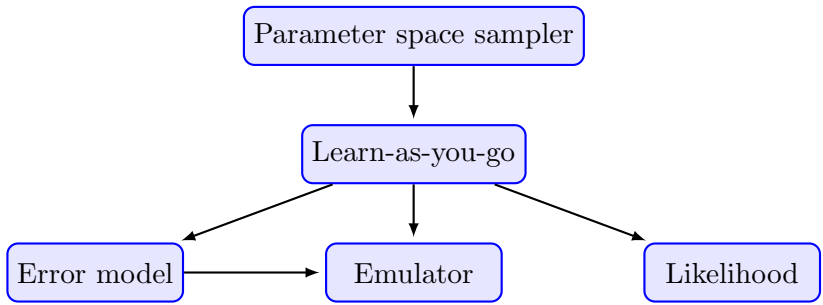

Cosmological parameter estimation using our algorithm has five major components:
\begin{enumerate}
  \item {\bf Sampler.---} An externally defined method to explore parameter space.
  \item {\bf Data likelihood.---}  The pre-defined function that we intend to emulate.
  \item {\bf Emulator.---}  Given a training set, perform a rapid emulation for new parameter values of interest.
  \item {\bf Error model.---}  For every new point, estimate the error made by emulating.  Propagate the errors in the likelihood from the emulation process into errors for the posterior distributions.
    \item {\bf Learn-as-you-go.---}  Continuously update the training set of the emulator based on new exact evaluations. Also, continuously update the error model.  Determine whether the emulated value is acceptable or if the exact calculation should be performed.
\end{enumerate}
Fig.~\ref{diagram_fig} describes the relationship between the different classes in our publicly available, object-oriented C++ implementation for parameter space sampling and likelihood evaluations.

Our techniques are not restricted  to cosmology and can be easily applied to any computationally expensive calculation that depends on a fixed number of input parameters. The algorithm does not make strong assumptions about the properties of the parameter space and  is usable with a wide variety of statistical samplers.
Frequently repeated operations take at most $O(\log \Ntr)$ time, where $\Ntr$ is the size of the training set, so that the speed of the algorithm is only weakly dependent on $\Ntr$.  Furthermore, the cost of evaluating the error model and adding new points to the training set is  negligibly small for typical calculations.
Our implementation scales well in an MPI environment, with all processes sharing a single training set.
Unlike the approach of Ref.~\cite{Bouland:2010mv} we do not assume that the parameter space is simply connected or has a single, global maximum.

Following Ref.~\cite{Bouland:2010mv}, we use the emulated value of the data likelihood for any set of parameters for which the emulation error is below a user-defined threshold; otherwise the exact calculation is performed and its result added to the training set.
Defining an unacceptable level of error between emulated and non-emulated values lets the user tradeoff between  accuracy and  acceleration, assuming that the error model is sufficiently robust.
This approach  ensures that there are no redundancies in the training set: new  points are not added where they are not needed and no time is spent generating training data that are not used.  In common with the techniques described in Refs~\cite{Bouland:2010mv,Graff:2011gv}, we do not need an initial training set, but providing one can significantly increase the resulting acceleration.

We demonstrate the use of the algorithm by accelerating estimates of posterior probabilities for $\Lambda$CDM with CMB likelihoods.  CMB analyses provide a useful testing platform for our methods because they can be conducted in a reasonable time but are still slow enough that any improvement in speed will be extremely welcome. Starting with an empty training set we use the emulator with two parameter space samplers: the Metropolis-Hastings \cite{mcmc} MCMC method  implemented in \textsc{Cosmo++}~\cite{Aslanyan:2013opa} and nested sampling with the \textsc{MultiNest} package~\cite{Feroz:2007kg,Feroz:2008xx,Feroz:2013hea}. We track the number of exact likelihood calculations required with and without emulation, as well as the overall speedup.  We demonstrate that the use of the emulator results in negligibly small errors in the posterior distributions of the parameters as determined by both our pre-defined error model and the exact posteriors obtained without emulation.

\section{Emulation Algorithm}\label{algorithm_sec}

    \begin{table}
      \centering
      \renewcommand{\arraystretch}{1.2}
      \setlength{\arraycolsep}{5pt}
      \begin{tabular}{ p{1.3cm} | p{5cm} || p{1.3cm} | p{5cm} }
        Symbol & Description &
        Symbol & Description \\
        \hline
        $p$               & Probability distribution function &
        $D$               & Data \\
        $\L$              & Data likelihood &
        $f$               & Exact function intended for emulation \\
        $\f$              & Emulating function that approximates $f$ &
        $n$               & Number of dimensions in parameter space (domain of $f$) \\
        $\x$              & Parameter space vector &
        $\Xtr$            & Training set of parameter space points \\
        $\Ytr$            & Range of training set $f(\Xtr)$ &
        $m$               & Number of dimensions for the range of $f$ \\
        $\rho$            & Model-dependent measure of local sparsity of training points &
        $k$               & Number of nearest neighbors used in analysis \\
        $w$               & Weighting of nearest neighbors &
        $C$               & Covariance matrix of training set \\
        $L$               & Cholesky decomposed covariance $C \equiv LL^T$ &
        $\delta$          & The degree of the polynomial interpolation for $\f$ \\
        $\df$             & Local emulation error, \emph{i.e.}, difference between $\f(\x)$ and $f(\x)$ &
        $e_\eta$          & Local emulation error on arbitrary scalar function $\eta$ \\
        $\Dp$             & Global emulation error in posterior probability &
        $\nuis$           & Nuisance parameters \\
        $r$ & Ratio of $e_\eta$ to $\rho$ \\
      \end{tabular}
      \caption{A legend for our notation.}
    \label{table:legend}
  \end{table}

  While cosmological applications will be primarily concerned with emulating the likelihood function $\L(D \| \x)$, for observational data $D$ and model parameters $\x$, in general we could emulate any arbitrary function $f(\x)$. In this Section we will keep our discussion as general as possible, and will introduce  an explicit local error model in Sect.~\ref{error_sec}.

Suppose we need to calculate the function
\begin{equation}
  \label{eqn:f}
  f:\mathbb{R}^n\rightarrow\mathbb{R}^m
\end{equation}
a large number of times with a specified  accuracy.
The exact numerical evaluation of $f$ is assumed to be slow, so we wish to store previously calculated results in order to rapidly estimate the value of $f$ for new input points $\x \in \mathbb R^n$.  The emulated value of $f$ is denoted by  $\f$.

A detailed schematic of the emulation algorithm is given in Alg.~\ref{alg:emulator} and
the implementation  of our publicly available code is discussed in Appendix \ref{app_impl}. In particular, the \kd tree used to  locate the points used for the emulation is described in   \ref{kdtree_app} and the emulator implementation is discussed in \ref{emulator_app}.
We expand on each of these functions below:

\begin{algorithm}[t]
  \caption{Emulates a function $f$ that maps $\x \in \mathbb R^n \to \y \in \mathbb R^m$, given a training set consisting of parameter space points $\{\x_i\}$ and their mappings $\{\y_i=f(\x_i)\}$.
}
  \label{alg:emulator}
  \begin{algorithmic}
    \Function{InitializeEmul}{$\Xtr=\{\x_i\,:\,i=\overline{1,\Ntr}\}$,$\Ytr=\{\y_i\,:\, i=\overline{1,\Ntr}\}$}
    \State Calculate the sample covariance matrix $C$ of points $\x_i$ in $\Xtr$
      \State Cholesky decompose $C$ into matrices $L$ by $C \equiv LL^T$
      \State Build training set in new basis $\Xtr^\prime = \{L^{-1}\x_i\,:\,i=\overline{1,\Ntr}\}$
      \State Build \kd tree $T_{kd}$ from $\Xtr^\prime$
    \EndFunction

    \Function{Emulate}{$\x_0$}
      \State Change basis by defining $\x_0^\prime \equiv L^{-1}\x_0$
      \State Using $T_{kd}$, find $k$ nearest neighbors $X^\mathrm{NN}=\{\x^\mathrm{NN}_i\,:\,i=\overline{1,k}\}$ of $\x_0^\prime$
      \State Find the corresponding $Y^\mathrm{NN} \equiv \{\y^\mathrm{NN}_i=f(\x_i^\mathrm{NN})\,:\,i=\overline{1,k}\}$ from $\Ytr$
      \ForAll{nearest neighbors $\x^\mathrm{NN}_i \in X^\mathrm{NN} $}
      \State Calculate weighting $w^i \equiv 1/\mathrm{Distance}(\x_0^\prime,\x^\mathrm{NN}_i)$ for some metric
      \EndFor
      \ForAll{$j=\overline{1,m}$}
      \State Assume $\left(f_\mathrm{em}(\x)\right)_j\equiv P_j(\x^\prime)$ around $\x_0^\prime$ where $P_j$ is a polynomial
    \State Find coefficients of $P_j$ to minimize $S_j=\sum_{i = 1}^{k} w^i |P_j(\x^\mathrm{NN}_i) - \left(\y^\mathrm{NN}_i\right)_j|^2$
      \EndFor
      \Return $f_\mathrm{em}(\x)=(P_1(\x_0^\prime),\dots,P_m(\x_0^\prime))$
    \EndFunction

    \Function{AddPointToTrainSet}{$\x$,$\y$}
      \State Add $\x$ to $\Xtr$ and $\y$ to $\Ytr$
      \State Add $\x^\prime \equiv L^{-1}\x$ to $\Xtr^\prime$ and $T_{kd}$
      \If{$\mathrm{Depth}(T_{kd})\geq4\log \Ntr$}
      \State Rebalance $T_{kd}$
      \EndIf
    \EndFunction

    \vspace{0.33cm}

      \Procedure{SimpleEmulator}{}
      \State  Calculate training sets $\Xtr$ and $\Ytr=f(\Xtr)$ exactly
      \State  Set up the emulator by calling \textsc{InitializeEmul}$(\Xtr,\Ytr)$
      \State  Emulate $f(\x)$ by calling \textsc{Emulate}$(\x)$
      \EndProcedure

  \end{algorithmic}
\end{algorithm}

\paragraph{\textsc{InitializeEmul}.---}  This function builds and saves a \kd tree from the training set $\Xtr$ following Ref.~\cite{Bentley:1975:MBS:361002.361007}.
We linearly transform the input points $\x \to \x^\prime$ for a new basis in which all of the parameters in the transformed training set are uncorrelated and where all of the components for each $\x^\prime$ have the same order of magnitude. To achieve this, we  calculate the  sample correlation matrix, $C$, of the points in the training set, find its Cholesky decomposition matrix $C=LL^T$, and use $L$ to change the basis to $\x^\prime \equiv L^{-1} \x$.

\paragraph{\textsc{Emulate}.---} This function computes $\f$ for an input point $\x_0$, finding the $k$ nearest neighbors ($k$-NN) of $\x$ in the \kd tree built from $\Xtr$.  These 
near-by points in the training set are then interpolated and the value of the polynomial interpolation at $\x_0$ is used for $\f(\x_0)$.

As the \kd tree in \textsc{InitializeEmul}  stores  input points in the Cholesky basis, $\x_0$ is first transformed to that basis with $\x_0 \to \x_0^\prime$, after which the $k$ nearest neighbors are located using the \kd tree. Given we are working with dimensionless, normalized variables we use the Euclidean distance to put a metric on the parameter space, but other possibilities exist including the Fisher information metric.  In general, $f(\x) = (f_1,\dots,f_m)$ with $m\ge1$, so for each dimension $j=1,\dots,m$ we build the $j^\mathrm{th}$ component of $\f$ by an independent, $n$-dimensional polynomial interpolation near the point $\x_0^\prime$. The coefficients of the polynomial are found simply by doing a weighted least squares fit to the $k$ nearest neighbors, with weights $w^i$ inversely proportional to the Euclidean distance. The details of the fit are discussed in Appendix \ref{app_interpolation}.

An $n$-dimensional polynomial of degree $\delta$ will have $N_\mathrm{coeff}$ coefficients, with
\begin{equation}
  \label{number_of_free_params}
  N_\mathrm{coeff} = 1 + n + \frac{n(n + 1)}{2} + \dots + \frac{n(n+1)\cdots(n+\delta - 1)}{\delta!}.
\end{equation}
To avoid degeneracies we use $k=2N_\mathrm{coeff}$. It is easy to see that $N_\mathrm{coeff}$ increases exponentially with $\delta$, so choosing a high dimensional polynomial will require a large number of nearest neighbors, as well as a least squares fit in a high-dimensional space, resulting in a uselessly slow emulator. Consequently, we have restricted our implementation to linear and quadratic polynomials.

\paragraph{\textsc{AddPointToTrainSet}.---} This function adds new points to the training set by first transforming to the Cholesky basis and  then adding them to the appropriate location in the \kd tree. After adding many  points  the \kd tree  may become unbalanced, resulting in a slowdown of the nearest neighbor search. After adding a new point we check if the tree has become  unbalanced, and rebuild it if necessary. The criterion for this step is that the depth of the tree is four times larger than the depth of a perfectly balanced tree.  This ensures that the tree  always has an $O(\log N_\mathrm{tr})$ depth. In practice, this rebuilding is sufficiently infrequent that it makes a negligible contribution to the overall computational cost of the algorithm.

\paragraph{\textsc{SimpleEmulator}.---}  This is the simplest possible implementation of the above functions.  After obtaining a training set, \textsc{InitializeEmul} builds the \kd tree with $\f(\x)$ computed by \textsc{Emulate} for a parameter space point $\x$.

\paragraph{Computational complexity.---}
If the dimensions $n$ and $m$ of the input and output spaces of $\f$, respectively, are much smaller than $N_\mathrm{tr}$,  the run-time of the algorithm  depends primarily on the size $\Ntr$ of the training set.
The complexity of \textsc{InitializeEmul} is dominated by the construction of the \kd tree, which is $O(\Ntr\log \Ntr)$. Finding the $k$ nearest neighbors takes only $O(\log \Ntr)$ time \cite{Bentley:1975:MBS:361002.361007} and the interpolation step is independent of $\Ntr$, which fixes the complexity of \textsc{Emulate}. Finally, adding a new point to the \kd tree is linear in the depth of the tree, \emph{i.e.}, adding new points is $O(\log \Ntr)$. However, the infrequent rebalancing step  scales as $O(\Ntr\log \Ntr)$.

\section{Error Analysis}\label{error_sec}

We emulate the function $f$ in Eq.~\eqref{eqn:f} with an approximate function $\f$ using the method described in Sect.~\ref{algorithm_sec}. However, we also want to estimate the difference between $f$ and $\f$.  This is critical both to the estimate of the total error in our calculation and when determining whether or not the approximation $\f(\x)$ is acceptable at a specific point $\x$.

We define the local emulation error $\df$ at a point $\x$ as
\begin{equation}
  \df(\x) \equiv \f(\x) - f(\x).
\end{equation}
Unless $\x$ is an element of the training set $\Xtr$ we will not know $f(\x)$, so we treat $\df$ as a random variable with a probability distribution $p(\df)$.
It is also convenient to define a local error $e_\eta$ via a  scalar function $\eta: \mathbb R^m \to \mathbb R$,
\begin{equation}
  e_\eta(\x) = \eta(\f(\x)) - \eta(f(\x)) \, .
\end{equation}
The form of $\eta$ is not unique, beyond requiring that it reduces the dimensionality of the output space of $f$ to unity.  Possible choices include a spatial average on the output space, $\eta(f(\x)) \equiv \vev{f(\x)}$, or by evaluating the likelihood for experimental data, $\eta(f(\x)) \equiv p(D \| f(\x))) $.\footnote{We will typically assume that $\eta(\y)$ is inexpensive to calculate compared to $f(\x)$, if we are given $\y$.  If it is not, then we would ideally emulate the composite function $f^\prime = \eta \circ f$ directly.}

We want an  error model that is broadly applicable, computationally efficient, and updates automatically as the training set grows.
We expect that $p(\df)$ and $p(e_\eta)$ will be strongly dependent on the distances between $\x$ and its nearest points in $\Xtr$, which are used for the interpolation.
We will use a single parameter $\rho(\x \| \Xtr)$ to define the probability distribution $p(e_\eta \| \rho, \Xtr)$.  We choose $\rho$ to be inversely proportional to the density of the training set $\Xtr$ near $\x$, typically the mean $n$-dimensional Euclidean distance to the $k$ nearest neighbors in $\Xtr$.

The form of $p(e_\eta \| \rho, \Xtr)$ could be modelled using \emph{a priori} knowledge of $f$ and $\eta$, but it is far more flexible to estimate $p(e_\eta\| \rho, \Xtr)$ empirically, via cross-validation on subsets $\Xcv$ of $\Xtr$, from which $e_\eta$ can be calculated exactly.  We assume that $e_\eta$ increases linearly with $\rho$,  allowing us to define a probability distribution on their ratio as
\begin{equation}\label{error_model_example}
  p(e_\eta \| \rho, \Xtr)  \rightarrow p\left(\frac{e_\eta}{\rho} \, \big | \, \Xcv \right),
\end{equation}
up to an arbitrary normalization.

\begin{algorithm}[t]
  \caption{Building an empirical model for the local error $e_\eta$ on the scalar function $\eta$, using a set of exact calculations $\Ycv = f(\Xcv)$ that the emulating function $\f$ has not yet been trained on.}
  \label{alg:error}
  \begin{algorithmic}
    \Function{BuildErrorModel}{$\eta$,$X_\mathrm{CV}=\{\x_i\,:\,i=\overline{1,M}\}$,$Y_\mathrm{CV}=\{\y_i=f(\x_i)\,:\,i=\overline{1,M}\}$}
      \State Initialize an empty list $R$
      \ForAll{points $\x_i$ in $ \Xcv$}
      \State Calculate $\rho(\x_i)$, \emph{e.g.}, as the mean distance to the $k$ nearest neighbors
      \State Add ratio $|\eta($\Call{Emulate}{$\x_i$}$) - \eta(\y_i)|/\rho(\x_i)$ to $R$, with \textsc{Emulate} from Alg.~\ref{alg:emulator}
      \EndFor
      \State Create histogram $H$ from $R$
      \State Smooth and normalize $H$ to obtain an estimate of $p(|e_\eta|/\rho)$
    \EndFunction

    \vspace{0.33cm}

    \Procedure{SimpleError}{}
    \State  Calculate sets $\Xcv$ and $\Ycv=f(\Xcv)$ exactly
    \State  Set up the emulator by calling \textsc{InitializeEmul}$(\Xtr,\Ytr)$
    \State  Estimate $p(|e_\eta|/\rho)$ by calling \textsc{BuildErrorModel}$(\eta,\Xcv,\Ycv)$
    \State  For point $\x$ calculate $p(e_\eta(\x)) \propto \rho(\x) \times p(|e_\eta|/\rho)$ and renormalize
    \EndProcedure

  \end{algorithmic}
\end{algorithm}

We describe our error model algorithm schematically in Alg.~\ref{alg:error} and its detailed implementation is given in Appendix~\ref{error_model_impl_app}.
The function \textsc{BuildErrorModel} takes as input the cross validation sets $\Xcv$ and $\Ycv = f(\Xcv)$.  For each point $\x_i$ in $\Xcv$ it calculates the local scalar error $e_\eta(\x_i)$ and the training set sparsity measure $\rho(\x_i)$, using the mean Euclidean distance to the $k$ nearest neighbors.  We enforce our assumption $e_\eta \propto \rho$ by constructing a smoothed and normalized histogram for  the ratio $|e_\eta|/\rho$, from which we empirically estimate $p(|e_\eta|/\rho)$.\footnote{We also save the $1\,\sigma$ upper bound $b_{1\sigma}$ on $p(|e_\eta| / \rho)$ for use in the learn-as-you-go algorithm.} \textsc{SimpleError} evaluates $p(e_\eta(\x))$ by  inverting the transformation in Eq.~\eqref{error_model_example}, multiplying $p(|e_\eta|/\rho)$ by $\rho(\x)$, allowing for $e_\eta$ to be positive or negative with equal probability, and renormalizing.

\section{Learn-As-You-Go Techniques}\label{learn_as_you_go_sec}

\begin{algorithm}[t]
  \caption{Implementation of emulation scheme (Alg.~\ref{alg:emulator}) and error model (Alg.~\ref{alg:error}) into a learn-as-you-go method that progressively builds a training set $\Xtr$, learns the error distribution $p(e_\eta)$, and decides when to emulate or exactly evaluate $f(\x)$.}\label{alg:learnasyougo}
  \begin{algorithmic}

    \Function{PrepareEmulator}{$\eta$,$\Xtr$,$\Ytr$,$\phi_\mathrm{CV}$}
      \If{$\mathrm{Size}(\Xtr)\geq N_\mathrm{min}$}
      \State Move a fraction $\phi_\mathrm{CV}$ of the points in $\Xtr$,$\Ytr$ to $X_\mathrm{CV}$, $Y_\mathrm{CV}$ (chosen randomly)
      \State Initialize the Emulator with \Call{InitializeEmul}{$\Xtr$,$\Ytr$}
      \State Build the Error Model with \Call{BuildErrorModel}{$\eta,X_\mathrm{CV}$,$Y_\mathrm{CV}$}
      \State Add $\Xcv$ and $\Ycv$ back to $\Xtr$ and $\Ytr$, respectively
      \State Save $\mathrm{Size}(\Xtr)$ as $\mathrm{CurrentSize}$
      \EndIf
    \EndFunction

    \Function{EmulationIsValid}{$\x$}
    	\State Calculate $\rho(\x)$, e.g., as the mean distance to the $k$ nearest neighbors
	\State Define $e_\eta^\mathrm{max} \equiv b_{1\sigma} \rho$ where $b_{1\sigma}$ is the $1\,\sigma$ upper bound of histogram $H$ from error model
      \State {\bf return } true {\bf if} $e_\eta^\mathrm{max}<\epsilon$, {\bf else return} false
    \EndFunction

    \Function{CalculateExact}{$\x$}
      \State Add $\x$ to $\Xtr$ and the exact value $\y=f(\x)$ to $\Ytr$
      \If{Emulator and Error model are initialized}
	\State Add $\x$, $\y$ to training set using \Call{AddPointToTrainSet}{$\x$,$\y$}
	\If{$\mathrm{Size}(\Xtr)\geq(1+\phi_\mathrm{ERR}) \times \mathrm{Current Size}$}
	  \State Reset the Emulator and the Error Model to uninitialized
	  \State Reinitialize them using \Call{PrepareEmulator}{$\eta$,$\Xtr$,$\Ytr$,$\phi_\mathrm{CV}$}
	\EndIf
      \Else
	\If{$\mathrm{Size}(\Xtr)\geq N_\mathrm{min}$}
	  \State Initialize using \Call{PrepareEmulator}{$\eta$,$\Xtr$,$\Ytr$,$\phi_\mathrm{CV}$}
	\EndIf
      \EndIf
      \State \Return{$\y$}
    \EndFunction

    \Function{CalculateLAYG}{$\x$}
      \If{Emulator and Error Model are initialized}
	\If{\Call{EmulationIsValid}{$\x$}}
	  \Return{\Call{Emulate}{$\x$}}
	\Else \,
	  \Return{$\y_\mathrm{exact} \equiv \textsc{CalculateExact}(\x)$}
	\EndIf
  \Else
  \State \Return{$\y_\mathrm{exact} \equiv \textsc{CalculateExact}(\x)$}
      \EndIf
    \EndFunction

    \vspace{0.33cm}

    \Procedure{LearnAsYouGo}{}
    \State  Define an external statistical sampler \textsc{StatSampler}, \emph{e.g.}, Metropolis-Hastings
    \For{$\x$ drawn from \textsc{StatSampler}}
    \State Evaluate and save the output from \Call{CalculateLAYG}{$\x$} as $\y$
    \EndFor
    \EndProcedure

  \end{algorithmic}
\end{algorithm}

We now combine the emulator from Sect.~\ref{algorithm_sec} and the error model from Sect.~\ref{error_sec}  into our learn-as-you-go algorithm.  We assume the use of an externally defined statistical sampler, \emph{e.g.}, Metropolis-Hastings or nested sampling.
The details of the emulation methods are completely independent of the sampling strategy. 
The resulting algorithm is outlined schematically in Alg.~\ref{alg:learnasyougo}; its detailed implementation is discussed in Appendix~\ref{layg_app}; and we describe the purpose of its separate functions below.

\paragraph{\textsc{PrepareEmulator}.---}

If the size of the training set $\Xtr$ exceeds a pre-defined threshold $N_\mathrm{min}$,  this function initialises  the emulator. \textsc{PrepareEmulator} is called multiple times: first when the training set size reaches $N_\mathrm{min}$ and then every time the training set size grows by a fraction $\phi_\mathrm{ERR}$, as explained below.

\textsc{PrepareEmulator} first divides the training set $\Xtr$ into a smaller training set $\Xtr^\prime \subset \Xtr$ and a cross validation set $\Xcv$ by randomly choosing a fraction $\phi_\mathrm{CV}$ of points from $\Xtr$ and removing them from the training set.  It is important to ensure that points in $\Xcv$ do not coincide with points in $\Xtr$, since points in $\Xtr$ have no emulation error by definition.  The remainder of the training set is temporarily used to initialize the emulator, while $\Xcv$ is used to initialize the error evaluation module.

After computing the error model from $\Xcv$, the points in this set are added back to $\Xtr$; further evaluations rely on the complete training set $\Xtr$.
Finally, \textsc{CurrentSize} stores the size of $\Xtr$ at the point when the emulator and the error model were last updated.  When the size of $\Xtr$ increases by the fraction $\phi_\mathrm{ERR}$ compared to \textsc{CurrentSize}, we update the emulator and error model.

\paragraph{\textsc{EmulationIsValid}.---}

This function determines whether or not the approximate value $\f(\x)$ should be accepted at a given point $\x$.  Following Alg.~\ref{alg:error}, we determine the value $b_{1\sigma}$ of the $1\,\sigma$ upper bound on the probability distribution function for the error ratio $p(|e_\eta|/\rho)$, for the scalar function $\eta$.  For a point $\x$,
\textsc{EmulationIsValid} calculates the average Euclidean distance $\rho(\x)$ to its $k$ nearest neighbors in $\Xtr$, then multiplies $b_{1\sigma}$ with $\rho(\x)$ to determine the $1 \sigma$ upper bound on $e_\eta(\x)$.  Emulation is accepted if this upper bound on the error,  $e_\eta^\mathrm{max}$, is less than a target error threshold $\epsilon$.

\paragraph{\textsc{CalculateLAYG} and \textsc{CalculateExact}.---}

As in  \textsc{LearnAsYouGo}, \textsc{CalculateLAYG} is called for any evaluation of $\f(\x)$.  This function first checks if the error in the emulation scheme is acceptable. If it is, then the emulator from Alg.~\ref{alg:emulator} is used to evaluate $\f(\x)$.
If the error is not acceptable, \textsc{CalculateExact} is called to exactly evaluate $f(\x)$ and add $\x$ to $\Xtr$ and $f(\x)$ to $\Ytr$.  Every time a new point is added to $\Xtr$, this function checks if the size of $\Xtr$ has grown by a fraction $\phi_\mathrm{ERR}$ since the error model was last  updated, in which case  \textsc{PrepareEmulator} is called.

\paragraph{\textsc{LearnAsYouGo}.---}

  Given a pre-defined method for obtaining points $\x$ from some probability distribution of interest, \textsc{CalculateLAYG} is called on $\x$ and its results are saved in a Markov chain.

\section{Propagating Local Errors to Posteriors}\label{posterior_error_sec}

Let us now specialize to the important case where the function $f(\x)$ in Eq.~\eqref{eqn:f}, which we intend to emulate, is the logarithm of a likelihood function
\begin{equation}
  f(\x) \to \ell(\x) \equiv \log \L(D \| \x),
\end{equation}
which is used by a statistical sampler to obtain posterior probability distributions.
The learn-as-you-go Alg.~\ref{alg:learnasyougo} allows for estimation of the local error $\Dl$ at each point $\x$, and we will translate this into a global uncertainty $\Dp(\x \| D)$ in the posterior probability distributions.
In this section we will explicitly calculate this error and in Sect.~\ref{cmb_sec} we will describe its use in cosmological parameter estimation problems.\footnote{Our publicly available code includes the posterior distribution error estimation, allowing the user to easily obtain the uncertainties on marginalized posterior distributions resulting from the emulation process.}

The posterior probability distribution of the parameter space vector $\x$ is related to the likelihood function $\mathcal{L}(D\|\x)$ by Bayes theorem:
\begin{equation}
  p(\x\|D)=\frac{1}{Z(D)} \, \mathcal{L}(D\|\x) \, p(\x),
  \label{eqn:bayes}
\end{equation}
where $p(\x)$ is the prior probability distribution for $\x$, and
\begin{equation}
  Z(D) \equiv \int\mathcal{L}(D\|\x) \, p(\x) \, d\x
\end{equation}
is  the Bayesian evidence or marginalized likelihood.

We will be interested in posterior distributions for an individual parameter  $x_0$, found after marginalizing over a set of nuisance parameters $\nuis$, given by
\begin{equation}\label{post_marg}
  p(x_0\|D)=\int p(x_0, \nuis \|D) \, d\nuis
  \;  = \; \frac{1}{Z}  \int \L(D\|x_0, \nuis) \, p(x_0, \nuis) \, d \nuis,
\end{equation}
where Bayes theorem~\eqref{eqn:bayes} was used in the second relationship.
The error in the emulated marginalized posterior
\begin{equation}
  \Dp(x_0 \| D) \equiv \pem(x_0 \| D) - p(x_0 \| D)
  \label{eqn:dp_def}
\end{equation}
can be expressed in terms of emulated quantities as
\begin{equation}
  \Dp (x_0 \| D) = \frac{1}{\Zem} \int p(x_0,\nuis) \, \DL (D \| x_0, \nuis) \, d \nuis - \left[ \frac{\pem (x_0 \| D)}{\Zem} \right]  \DZ(D)
  \label{eqn:Dp}
\end{equation}
to first-order in the error terms $\DL$ and $\DZ$,
where
\begin{equation}
  \DZ(D) = \int p(\x)  \,\DL (D \| \x) \, d \x.
  \label{eqn:DZ}
\end{equation}
The error in the Bayesian evidence $\Delta Z$ and data likelihood $\DL$ are defined similarly to Eq.~\eqref{eqn:dp_def}, where emulated quantities
are evaluated using $\lem$ from the learn-as-you-go emulation scheme of Sect.~\ref{learn_as_you_go_sec}.

Assuming  the errors $\Dp$ and $\DL$ are small, we can rearrange Eqs~\eqref{eqn:Dp}~and~\eqref{eqn:DZ} as
\begin{eqnarray}
  \label{eqn:post_error}
  \Delta \log p (x_0 \| D) &=& \int \Delta \log \L (D \| x_0, \nuis) \, \mathcal P (\nuis \| x_0) \, d \nuis\\
                           &-& \int \Delta \log \L (D \| x, \nuis)  \, \pem (x, \nuis \| D) \, d x \, d\nuis,
  \notag
\end{eqnarray}
where we have defined the normalized probability distribution
\begin{equation}
  \mathcal P (\nuis \| x_0) = \frac{\pem (x_0, \nuis \| D)}{ \pem (x_0 \| D)}.
\end{equation}
The first term in Eq.~\eqref{eqn:post_error} is the contribution to the posterior error from the local region around $x_0$ and the second term is the uncertainty in the overall normalization for $p(x_0 \| D)$.  The posterior error $\Delta p(x_0 \| D)$ is proportional to the posterior itself, making the emulation error greatest near regions of large \emph{a posteriori} probability.

The learn-as-you-go emulation algorithm in Sect.~\ref{learn_as_you_go_sec} yields a sample of points $\Xsamp$ that are distributed according to $\pem (\x \| D)$, which we will use to evaluate Eq.~\eqref{eqn:post_error}.
Given $\Xsamp$ we first calculate $\pem(x_0 \| D)$ by histogram estimation, marginalizing over all other parameters and taking only the $x_0$ component of each element $\x \in \Xsamp$.
Specifically, by taking only the $x_0$ dimension from the elements of $\Xsamp$ and dividing the result into $N_\mathrm{bins}$ bins, we can approximate $p(x_0 \| D)$ by the fraction of the total sample that is in the bin to which $x_0$ belongs, with the global normalization set by requiring $\int p(x_0 \| D) dx_0 \equiv 1$.

Since any $x_0^\prime$  in the same bin as $x_0$ will have the same emulated posterior probability, we can obtain a sample of nuisance parameters $\Theta_\mathrm{samp}$ that follow $\mathcal P(\nuis \| x_0)$ by taking the $\nuis$ from those points $\x^\prime \in \Xsamp$ that have their $x_0^\prime$ in the same bin as $x_0$.
We then estimate the integrals in Eq.~\eqref{eqn:post_error} by
\begin{equation}
  \Delta \log p(x_0 \| D) \approx \frac{1}{N_\mathrm{bin}} \sum_{i=1}^{N_\mathrm{bin}} \Dl (D \| x_0, \nuis^i )
  - \frac{1}{\Nsamp} \sum_{j=1}^{\Nsamp} \Dl (D \| \x_j)
  \label{eqn:dp_sum}
\end{equation}
for $\x_j \in \Xsamp$ and $\nuis^i \in \Theta_\mathrm{samp}$.

With a probabilistic error model for $\Dl(\x)$ we can use the Lyapunov central limit theorem \cite{clt} to conclude that the result of each of the sums in Eq.~\eqref{eqn:dp_sum} will be normally distributed, assuming that the $\Dl$ are uncorrelated for different points and both $N_\mathrm{bin}$ and $\Nsamp$ are sufficiently large.
We expect the correlations in the $\Dl$ to be small in our learn-as-you-go emulation scheme, since the emulator is only used for points where the training set is dense.  Consequently, two nearby points will typically have different interpolation functions over their nearest neighbors, resulting in an essentially independent error.
For each emulated point $\x$ in $\Xsamp$, Alg.~\ref{alg:error} will calculate the mean $\mu_\ell(\x)$ and variance $\sigma^2_\ell(\x)$ of the probability distribution for $\Dl(\x)$.  For each point that is exactly evaluated, $\Dl (\x) \equiv 0$.

Finally, the probability distribution for the error in the log-posterior approaches
\begin{equation}
  p\left(\Delta \log p (x_0 \| D) \right) \; \xrightarrow{\mathrm{CLT}} \; \mathcal N \left[ \bar \mu_0, \bar \sigma_0 \right],
  \label{eqn:dp_gauss}
\end{equation}
where $\mathcal N$ is a normal distribution with mean
\begin{equation}
  \bar \mu_0 \equiv \frac{1}{N_\mathrm{bin}} \sum_{i=1}^{N_\mathrm{bin}} \mu_\ell (x_0, \nuis^i) - \frac{1}{\Nsamp} \sum_{j=1}^{\Nsamp} \mu_\ell(\x_j)
\end{equation}
and variance
\begin{equation}
  \bar \sigma_0^2 \equiv \frac{1}{N_\mathrm{bin}^2} \sum_{i=1}^{N_\mathrm{bin}} \sigma_\ell^2 (x_0, \nuis^i) + \frac{1}{\Nsamp^2} \sum_{j=1}^{\Nsamp} \sigma_\ell^2(\x_j).
  \label{eqn:dp_sigma}
\end{equation}
We have defined the error model in Sect.~\ref{error_sec} so that $\mu_\ell =0$, which makes $\bar \mu_0$ vanish.
This gives a simple way to evaluate the error in each bin of the marginalized posterior $\pem(x_0 \| D)$ by  combining the variances of the $\Dl$ for each point in the sample.
By defining an upper limit $\sigma_{\mathrm{max},\ell}^2$ on the allowed acceptable variance in the error model for $\Delta \ell$, Eq.~\eqref{eqn:dp_sigma} bounds the variance in the error in the log-posterior as $\sigma_0^2 \lesssim \sigma_{\mathrm{max},\ell}^2/N_\mathrm{bin}$, assuming $ N_\mathrm{bin} \ll \Nsamp$.

In practice we first evaluate the posterior probability distribution $\pem(x_0 \| D)$ from $\Xsamp$ by histogram estimation, as well as the local error in each bin, \emph{i.e.}, the first term in Eq.~\eqref{eqn:dp_sum}, ignoring the issue of normalization. Once the error is evaluated in each bin, we calculate the total normalization factor by summing all the bins and adding the errors of each bin according to the central limit theorem.  We then divide each bin by the normalization factor. Keeping only first-order terms, we add the error coming from normalization to the error in each bin, to estimate the total error.

\section{CMB and \emph{Planck} Likelihood}\label{cmb_sec}

We now apply this algorithm to CMB power spectrum and \emph{Planck} likelihood calculations, using the 2013 likelihood code~\cite{Ade:2013kta}.\footnote{Our public code can be easily used with the new \emph{Planck} likelihood code when it becomes available.}  We show posterior probabilities for both emulated and non-emulated scenarios in Figs~\ref{post_mcmc_fig}--\ref{post_combo_fig}, including the Gaussian-distributed posterior error as calculated in Eq.~\eqref{eqn:dp_gauss}.  We compare different values of the tolerance on the cutoff for the allowed error in the log-likelihood as well as two standard sampling techniques to show the robustness of this approach. Finally, we discuss the speed-ups achieved.

\subsection{Details of procedure}

The \emph{Planck} likelihood function is a product of several distinct independent likelihoods, the low-$l$ (\textsc{Commander}) and high-$l$ (\textsc{CamSpec}) temperature likelihoods, the lensing likelihood, and the WMAP polarization likelihood. \textsc{Commander} depends on the cosmological model through the CMB $TT$ power spectrum only, while \textsc{CamSpec} is also a function of $14$ nuisance parameters. The lensing likelihood takes as an input the $TT$ and $\Phi\Phi$ power spectra for the model, where $\Phi$ is the lensing potential, and the WMAP polarization likelihood takes the $TT$, $EE$, $TE$, and $BB$ spectra of the model as its input.

The \textsc{CamSpec} likelihood can be evaluated relatively quickly and is the only part of the total \emph{Planck} likelihood that depends on the additional nuisance parameters, so we exclude it from the emulation scheme to significantly reduce the dimensionality of the training set. Our approximation algorithm thus uses only the cosmological parameters as input parameters and  provides the $C_l^{TT}$, and \textsc{Commander}, lensing, and polarization likelihoods.  Each set of cosmological parameters will be calculated using the learn-as-you-go approximation algorithm, which will decide whether or not to use the exact numerical solution for the outputs. \textsc{CamSpec} is  called with the interpolated $C_l^{TT}$ values and the foreground parameters, completing the likelihood calculation.

The number of cosmological parameters $n$ depends on the cosmological model, but typically $n \sim \mathcal O(10)$.\footnote{For $\Lambda$CDM the number of parameters is $6$.} From Eq.~\ref{number_of_free_params} we choose quadratic interpolation between the nearest neighbors, for which we need $1+n+n(n+1)/2$ free parameters.

We use the simple error model described in Sect.~\ref{error_sec}. We set $\phi_\mathrm{ERR}=0.05$, which uses $5\%$ of the training set as a cross-validation set. However, when the cross-validation set size becomes larger than $1,000$ we keep only $1,000$ points. The function $\eta$ chosen for error evaluation is the entire \emph{Planck} likelihood function, evaluated from the output of the emulator. Specifically, the emulator calculates $C_l^{TT}$ values, as well as the \textsc{Commander}, lensing, and polarization likelihoods. The function $\eta$ passes the $C_l^{TT}$ and the best-fit values of the foreground parameters to \textsc{CamSpec}.  The result of this is combined with the other likelihoods. We choose the weighted average distance to the nearest neighbors as the measure $\rho$ of the local sparsity of the training points.

The acceptable amount of error in $\Dl$ is defined by the ratio $r_\mathrm{2 \sigma}$ at which the cumulative distribution equals $0.955$.  For each  set of input parameters we then find its $k$ nearest neighbors in $\Xtr$, the weighted average Euclidean distance to them, and multiply the distance by $r_\mathrm{2 \sigma}$ to find the $2\,\sigma$ upper bound on the error at that point.  We choose a threshold of $0.4$ for the $2\,\sigma$ upper bound on the error of $-2\log \mathcal{L}$, which corresponds to a $0.1$ threshold for the $1\,\sigma$ upper bound on $\log \mathcal{L}$ for Gaussian errors.  We use the emulator only if the $2 \sigma$ error at a point is below this threshold; otherwise the likelihood is calculated exactly.  Emulation  is not attempted until there are at least $N_\mathrm{min}=10,000$ points in the training set.

The training set is continuously updated throughout the run, with each new point being immediately added to a \kd tree.  We re-balance the tree as needed, using the criterion described in Sect.~\ref{algorithm_sec}.  We update the error model every time the training set increases in size by $25\%$, since likelihood evaluations for points in the cross-validation set are relatively time consuming.

We use a parallel implementation in which all threads share their training sets each time a single thread has accumulated  $10$ new points.  When using many independent Markov chains this gives  a significant performance boost, since chains that start out in different regions of the  parameter space soon explore regions  other chains have visited. Furthermore, we find that we  frequently sample the same regions of cosmological parameter space with different values of the foreground parameters. Since these are used in the relatively fast \textsc{CamSpec} likelihood module, we have effectively implemented a fast-slow split analogous to that provided by \textsc{CosmoMC}~\cite{Lewis:2002ah}, but without modifying the sampler itself.

\subsection{Emulated $\Lambda$CDM posteriors}
\label{ssect:lcdm_post}

\begin{figure}
\centering
\includegraphics[width=1.0\textwidth]{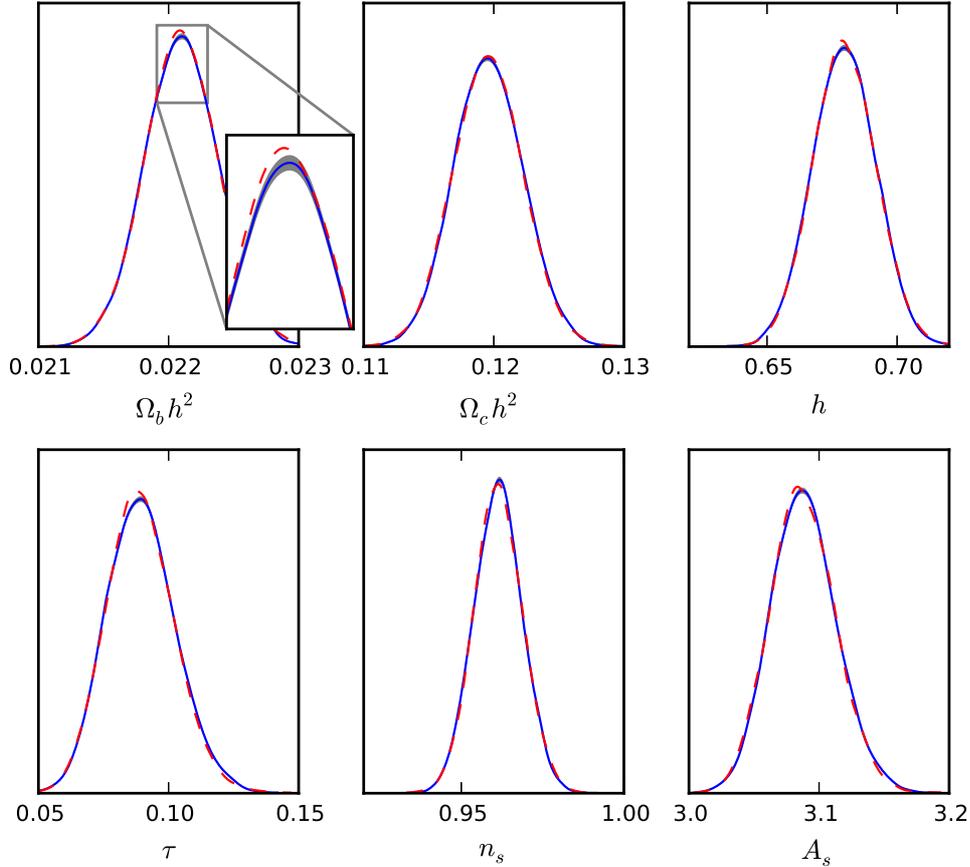}
\caption{\label{post_mcmc_fig} Posterior distributions of the cosmological parameters from MCMC sampling with (solid blue lines) and without (dashed red lines) emulation. The $1\,\sigma$ error of the posteriors for the case of emulation is shown as a gray band around the solid blue lines.}
\end{figure}

\begin{figure}
\centering
\includegraphics[width=1.0\textwidth]{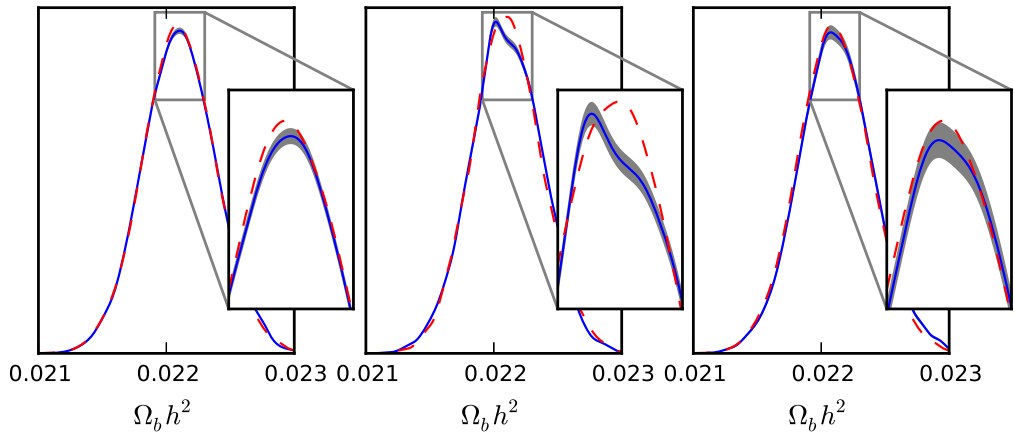}
\caption{\label{post_combo_fig} Posterior distribution of $\Omega_bh^2$ with (solid blue lines) and without (dashed red lines) using emulation. The $1\,\sigma$ error of the posteriors for the case of emulation is shown as a gray band around the solid blue lines. The left panel is for MCMC (identical to the top left plot in Fig.~\ref{post_mcmc_fig}, the middle panel is for \textsc{MultiNest}, and the right panel is for MCMC again but with a larger error threshold (see the discussion in the text).}
\end{figure}

\begin{table}
\renewcommand{\arraystretch}{1.5}
\setlength{\arraycolsep}{5pt}
\begin{eqnarray*}
\begin{array}{c|c|c|c|c}
  \text{Param.} & \text{MCMC exact} & \text{MCMC approx.} & \textsc{MultiNest} \text{ exact} & \textsc{MultiNest} \text{ approx.} \\
\hline
  \Omega_bh^2 & 0.02210\pm0.00028 & 0.02210\pm0.00029 & 0.02210\pm0.00028 & 0.02210\pm0.00028 \\
  \Omega_ch^2 & 0.1196\pm0.0027 & 0.1196\pm0.0027 & 0.1195\pm0.0026 & 0.1195\pm0.0027 \\
  h & 0.680\pm0.012 & 0.680\pm0.012 & 0.0680\pm0.012 & 0.0680\pm0.012 \\
  \tau & 0.089\pm0.013 & 0.089\pm0.013 & 0.089\pm0.013 & 0.089\pm0.013 \\
  n_s & 0.9613\pm0.0074 & 0.9614\pm0.0075 & 0.9613\pm0.0071 & 0.9615\pm0.0074 \\
  A_s & 3.087\pm0.025 & 3.088\pm0.025 & 3.087\pm0.026 & 3.088\pm0.025 \\
  \log Z & - & - & -4944.0\pm0.3 & -4944.1\pm0.3 \\
\end{array}
\end{eqnarray*}
\caption{Parameter constraints from MCMC and \textsc{MultiNest} samplers with and without using the emulator.}
\label{results_table}
\end{table}

In Fig.~\ref{post_mcmc_fig} we plot the posterior distributions for the cosmological parameters of the standard $\Lambda$CDM model using the \emph{Planck} 2013 temperature data together with WMAP polarization.  We use the  \textsc{Cosmo++} MCMC sampler and compare the posterior distributions of the cosmological parameters $\Omega_bh^2$, $\Omega_ch^2$, $h$, $\tau$, $n_s$, and $A_s$ with and without emulation. We also performed this comparison for the \textsc{MultiNest} sampler, with essentially identical results.  Fig.~\ref{post_combo_fig}  shows the posterior of one parameter, $\Omega_bh^2$ obtained with MCMC (left panel) and \textsc{MultiNest} (middle panel).  In all cases we start with an initially empty training set.

We  plot the $1 \sigma$ error range on the posteriors using the method in Sect.~\ref{posterior_error_sec} with gray bands around the solid blue lines  corresponding to the mean prediction. However, the errors are negligible and only visible if we magnify the axes substantially. Note that all posteriors include additional errors  resulting from the finite sizes of the samples used to estimate the distributions, which are not included in our estimation of the errors.  The uncertainty in the posterior induced by the emulation algorithm is smaller than this sampling uncertainty;  and the posteriors with and without emulation agree very well.

The right panel of Fig.~\ref{post_combo_fig} demonstrates a more substantial error on the posterior distributions by repeating the MCMC run with only two chains and a higher threshold for the error. We have allowed points to be emulated whenever the $2\,\sigma$ upper bound on $-2 \log \mathcal L$ is less than $1.0$ instead of $0.4$. Using only two chains compared to $10$ decreases the number of points in the sample by a factor of $5$, and increases the error bars by a factor of $\sqrt{5}$ (see Eq.~(\ref{eqn:dp_sigma})).  The error band for the posterior is wider than in the previous cases (left and middle panels) but is still comparable to the errors coming from the finite size of the sample.

\begin{figure}
\centering
\includegraphics[width=0.75\textwidth]{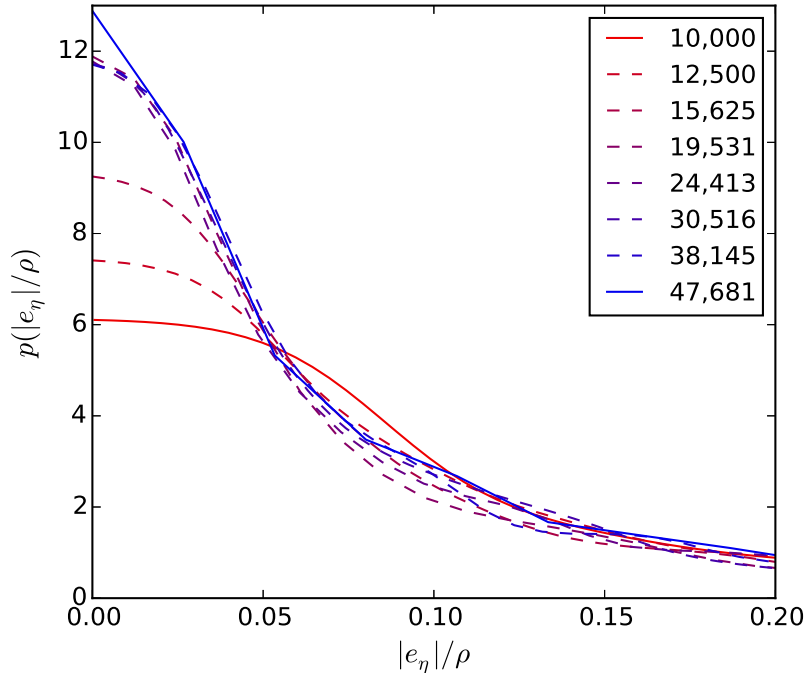}
\caption{\label{ratios_fig} Probability distribution of $|e_\eta|/\rho$ for different stages of error model evaluation (see Section \ref{error_sec}) for the MCMC run. The labels show the size of the training set.}
\end{figure}

The $1\,\sigma$ confidence intervals for the cosmological parameters and  Bayesian evidences\footnote{Computed using \textsc{MultiNest}.} are summarized in Table~\ref{results_table}. The results with emulation are almost identical to the exact case. Specifically, the confidence interval widths for the parameters vary by less than $5\%$, and the central values vary by less than $\sigma / 20$. Such small numerical differences can arise even if two runs are performed with the exact same likelihood, so the emulated likelihoods involve no sacrifice of accuracy with this set of run parameters.

Figure~\ref{ratios_fig} shows the estimated probability distribution of the errors $p(|e_\eta|/\rho)$ for all the stages of error evaluation during the MCMC run.
The error model is periodically updated every time the training set size increases by a fraction $\phi_\mathrm{ERR}$, which we have chosen to be $25\%$.
The red solid line shows the first evaluation, the blue solid line shows the last one, and the dashed lines show the intermediate stages. The increase of the training set size results in distributions with smaller expected variance for $p(|e_\eta|/\rho)$ and a progressively  higher peak at $e_\eta/\rho = 0$.

\subsection{Estimated improvement}
\label{ssect:speedup}

\begin{table}
\renewcommand{\arraystretch}{1.5}
\setlength{\arraycolsep}{5pt}
\begin{eqnarray*}
\begin{array}{c|c|c}
  \text{Calculation step} & \text{Linux + intel (16 cores)} & \text{Darwin + GNU (4 cores)} \\
\hline
\text{CMB} & 0.847\,\mathrm{sec} & 2.703\,\mathrm{sec} \\
\text{\textsc{CamSpec}} & 0.008\,\mathrm{sec} & 0.044\,\mathrm{sec} \\
\text{\textsc{Commander}} & 0.062\,\mathrm{sec} & 0.167\,\mathrm{sec} \\
\text{Polarization} & 0.190\,\mathrm{sec} & 1.035\,\mathrm{sec} \\
\text{Lensing} & 0.320\,\mathrm{sec} & 1.005\,\mathrm{sec} \\
\text{Rapid likelihood} & 0.014\,\mathrm{sec} & 0.050\,\mathrm{sec} \\
\hline
\multicolumn{3}{c}{\text{Speedup factors}} \\
\hline
\text{Temperature only} & 67 & 57 \\
\text{Temp. + pol.} & 81 & 78 \\
\text{Temp. + lens.} & 91 & 78 \\
\text{Temp. + pol. + lens.} & 105 & 99 \\
\end{array}
\end{eqnarray*}
\caption{The average times of different steps of likelihood evaluation, determined from $100$ repeated calculations (top). The speedup factors by using the emulator with $150,000$ training points (bottom).}
\label{speedup_table}
\end{table}

In Table \ref{speedup_table} we have estimated the speedup from using the emulator for a single likelihood calculation with a well-developed training set of $150,000$ points. We perform $100$ exact calculations of the CMB power spectra and \emph{Planck} likelihoods to evaluate the average run-time for each step and compare this to $100$ rapid calculations using the emulator. We estimate the run-times on two different architectures, Darwin with $4$ CPU cores, and Linux with $16$ CPU cores, using GNU compilers on Darwin and Intel compilers with the MKL library on Linux.

The slowest calculation is the CMB power spectrum.  The polarization and lensing likelihood calculations are roughly as fast as the CMB power spectra calculation; \textsc{Commander} and \textsc{CamSpec} are approximately one and two orders of magnitude faster, respectively. The emulated likelihood approximation is only slightly slower than \textsc{CamSpec}, meaning that the emulation itself takes a negligible amount of the run-time.  Furthermore, since the computational complexity depends only logarithmically on the training set size $\Ntr$, changing $\Ntr$ will have a negligible impact on the overall speed.

The speedups shown in Table \ref{speedup_table} are only achievable after obtaining a large training set. However, in the learn-as-you-go scenario we start with an empty training set with the algorithm training itself with exact likelihood evaluations.  Gradually the estimated error in the emulator will decrease with an increasing size for the training set and the estimated likelihood values will be reliable for new points. We will now estimate the overall speedup that can be gained using the learn-as-you-go approach separately for the MCMC and \textsc{MultiNest} samplers.

We start both samplers with an empty training set and compare the results with the case of no emulation. We run MCMC using adaptive posteriors with $10$ parallel chains. The burn-in length is chosen to be $500$ and the Gelman-Rubin convergence diagnostic~\cite{Gelman:1992ts,Aslanyan:2013opa} is used  with a convergence criterion of $1.01$. Convergence occurs when the chains reach a length of approximately $6,000$ in both cases, with or without emulation. When no emulation is used and all of the processes are run on nodes of similar speeds the resulting chains have similar sizes. However, when using the emulator, different chains may progress at significantly different speeds depending on how frequently they walk into dense regions of the training set, resulting in chains of significantly different lengths.  Convergence in this case is determined by the size of the slowest chain.
\textsc{MultiNest} can also be parallelized very well, so we run it with $10$ MPI processes and the recommended values of the other parameters for accurate evidence calculation~\cite{Feroz:2008xx}.

\begin{figure}
\centering
\includegraphics[width=0.49\textwidth]{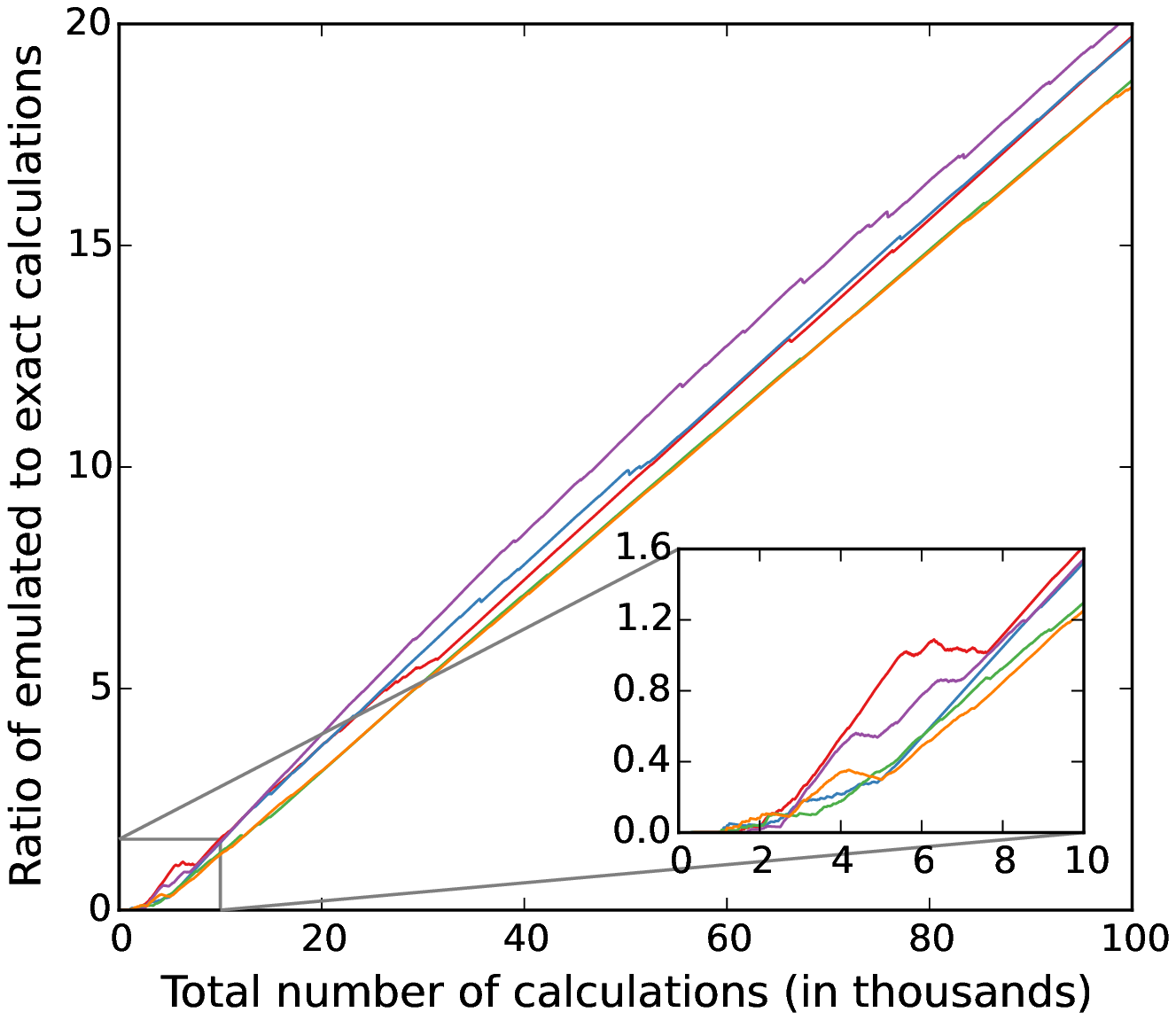}
\includegraphics[width=0.49\textwidth]{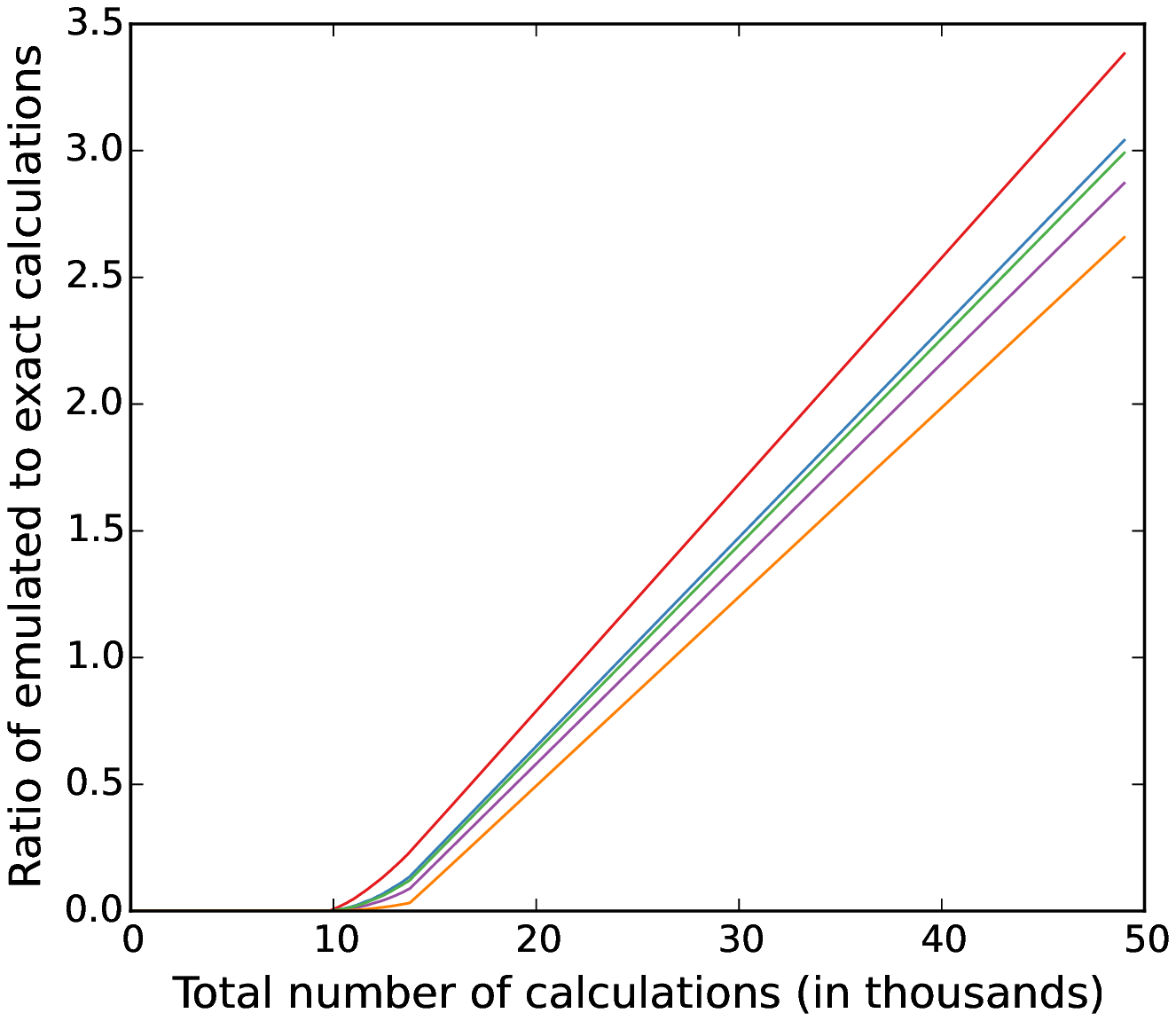}
\caption{\label{counts_mcmc_fig} The ratio of the number of emulated to the number of exact likelihood evaluations as a function of the total number for the case of MCMC sampling (left) and \textsc{MultiNest} sampling (right). Different colors represent $5$ randomly selected chains.}
\end{figure}

We plot the ratio of the number of emulated to the number of exact likelihood calculations in Fig. \ref{counts_mcmc_fig}.  The left panel has $5$ selected MCMC chains and the right panel has $5$ MPI processes for nested sampling.  The different MCMC chains have significantly different lengths so we cut off all of the lines to match the smallest chain.

These plots illustrate important characteristics of our learn-as-you-go approach. At the beginning of the run the ratio of emulated points starts and remains close to zero as the emulator  builds a training set.
However, once the emulator achieves a high-enough accuracy the ratio of emulated points increases roughly linearly.  This divides the run into ``mainly training'' and ``mainly emulation'' periods for each chain, with a transition occuring after about $5,000$ total evaluations per chain for the MCMC case and after about $10,000$ for \textsc{MultiNest}.
In the ``mainly emulation'' period, the ratio of emulated to exact calculations depends linearly on the total number of evaluations with a slope of $\alpha \approx 2.11\e{-4}$ for MCMC and $\alpha \approx 7.5\e{-5}$ for nested sampling.
With a slope of $\alpha$, the total number of exact evaluations  per chain $N_\mathrm{ex}$ will eventually saturate at
$N_\mathrm{ex} = 1/\alpha$ as the total number of evaluations gets large, which is approximately $4.8\e{3}$ and $1.3\e{4}$ for MCMC and nested sampling, respectively.  These upper limits are nearly coincident with the end of the ``mainly training'' period, indicating that there are almost no exactly evaluated calculations required during the ``mainly emulation'' stage.

\textsc{MultiNest} implements nested sampling, so it starts in regions of small likelihoods and gradually converges to the peak of the likelihood \cite{Feroz:2007kg,Feroz:2008xx,Feroz:2013hea}. This means that the training set is initially spread  thinly across the parameter space, making emulation impossible. However, \textsc{MultiNest} eventually converges on the peak of the likelihood, where the training set becomes sufficiently dense. The nested sampler then remains in the high likelihood regions, resulting in a more abrupt transition from ``mainly training'' to ``mainly emulation'' periods for \textsc{MultiNest} than for MCMC, as can be seen in Fig.~\ref{counts_mcmc_fig}.

For MCMC sampling, the ratio of emulated to exact calculations is approximately $19$ at the end of the run for the worst case chain and emulation was performed for about $95\%$ of the points. Since the speedup factor for the temperature plus polarization likelihoods is about $80$ in Table~\ref{speedup_table}, an overall speedup factor of about $16$ would be expected if there were no other calculations performed.
In practice, the wall-clock run-times for the  MCMC calculations improved by a factor of $6.5$ when run with emulation and an initially empty training set.
For \textsc{MultiNest}, Fig. \ref{counts_mcmc_fig} shows that the final ratio of the number of emulated to the number of exact calculations reaches about $2.5$--$3.5$ for different processes, compared to $19$--$20$ for the MCMC sampler. For all the $10$ chains, emulation was performed for $75\%$ of all likelihood evaluations. Without any computational overhead, this would give an expected speedup of $3.9$; in practice we obtained a value of $3.5$.

\section{Conclusion}\label{summary_sec}

We have incorporated an explicit error model in a flexible learn-as-you-go emulation algorithm.  When the slow calculation $f(\x)$ is a data likelihood function we have also shown that the estimated local error in the log-likelihood $\ell$ can be easily propagated into estimates of a model's posterior probability distributions.  Using a trustworthy error model  then removes any need for performing redundant and slow comparisons to exact calculations.

In Sect.~\ref{algorithm_sec} we described an emulation algorithm that approximates an arbitrary function $f: \mathbb R^n \to \mathbb R^m$ at a parameter space point $\x$ by using polynomial interpolation over the $k$ nearest neighbors to $\x$ in a pre-computed training set $\Xtr$.  In Sect.~\ref{error_sec} we implemented a simple error model to approximate the local difference between $f(\x)$ and its emulation $\f(\x)$.  Given an arbitrary measure $\rho$, which is inversely proportional to the density of the training set $\Xtr$ at the point $\x$, our model requires the error to scale like $\df(\x)\sim \rho$.  We have chosen $\rho$ to be the weighted mean Euclidean distance to the $k$ nearest training points, although this could be  relaxed in a more general error model.  We empirically estimate a probability distribution $p(\df(\x))$ by cross-validation on a subset $\Xcv$ of the pre-computed training set $\Xtr$.  Sect.~\ref{posterior_error_sec} specializes to the case where we emulate a cosmological likelihood function $\L$ and presents a method for propagating the local error $\DL$ to the error in marginalized posterior probability distributions.  Since the error model is probabilistic, we can then report a probability distribution for the value of the posterior probability $p(x_0 \| D)$, which tends to a normal distribution by the central limit theorem.

Improving on Ref.~\cite{Bouland:2010mv}, Sect.~\ref{learn_as_you_go_sec} implements a pre-defined tolerance $\epsilon$ on the allowed error $\df(\x)$ in the emulated function.  Parameter values that have emulation errors with a significant probability to be above a pre-defined threshold are directly computed, instead of emulated.  Points that are calculated exactly are added to the training set $\Xtr$ and the error model is updated repeatedly based on the increasing size and range of $\Xtr$.  By adjusting this threshold, the algorithm emulates more or less liberally, with a corresponding trade-off between run-time improvement and accuracy.
The specific details of the emulation algorithm, the error model, and the learn-as-you-go scheme can be understood from Algs~\ref{alg:emulator}, \ref{alg:error}, and \ref{alg:learnasyougo}, respectively.

We have applied this methodology to $\Lambda$CDM parameter estimation from CMB likelihoods in Sect.~\ref{cmb_sec}.  We chose CMB likelihoods because they are sufficiently slow to calculate that we see a moderate speed improvement when emulating, but are fast enough that we can exactly evaluate a significant fraction of the CMB Markov chains to develop a dense training set.  Section~\ref{ssect:lcdm_post} and Figs~\ref{post_mcmc_fig}--\ref{post_combo_fig} have shown that the error in the emulation scheme is sub-dominant to the finite-sample statistical error when we choose typical cutoff values on the probability distribution of the error in  $-2 \log \mathcal L$.  Section~\ref{ssect:speedup} approximates a speed-up of $\mathcal O(3-10)$ without the use of an initial training set with either MCMC or nested sampling methods.
The learn-as-you-go algorithm that we have developed is parallelized, fast, and applicable to general situations outside of CMB likelihood emulation.
We have publicly released all of our code in the \textsc{Cosmo++} package and have included the implementation details in Appendices~\ref{app_interpolation}~and~\ref{app_impl}.

\acknowledgments

The authors acknowledge the contribution of the NeSI high-performance computing facilities. New Zealand's national facilities are provided by the New Zealand eScience Infrastructure (NeSI) and funded jointly by NeSI's collaborator institutions and through the Ministry of Business, Innovation \& Employment's Research Infrastructure programme.\footnote{\url{http://www.nesi.org.nz}}

\appendix

\section{Least Squares Fit}\label{app_interpolation}

This appendix discusses the details of the weighted least squares fit of Alg.~\ref{alg:emulator} in Sect.~\ref{algorithm_sec}, which is performed by mimimizing the sum
\begin{equation}
S_j=\sum_{i = 1}^{k} w^i |P_j(\x^\mathrm{NN}_i) - \left(\y^\mathrm{NN}_i\right)_j|^2\,.
\end{equation}
The index $j$ denotes the output coordinate of the target function $f$ and the index $i$ runs over all the neareast neighbors.  The $j$-th component of $\y^\mathrm{NN}_i$ is $\left(\y^\mathrm{NN}_i\right)_j$. The weights $w^i$ are inversely proportional to the distances of the query point $\x_0^\prime$ to the nearest neighbors $\x^\mathrm{NN}_i$.

For convenience, we choose the polynomial to have the following form:
\begin{equation}
  P_j(\x)=a_j + \sum_{l = 1}^{n}b_j^l\left(\x -\x_0^\prime\right)^l+ \dots + \mathrm{terms}\;\mathrm{of}\;\mathrm{degree}\;\delta\,,
\end{equation}
where the upper index $l$ denotes the $l$-th component of a vector. Since we will only need to evaluate $P_j$ for $\x_0^\prime$, we will only need to calculate the value of $a_j$.

The coefficients of the polynomial $P_j$ can be written as a vector
\begin{equation}
\mathbf{p}_j=(a_j,b_j^1,\dots,b_j^n,\dots,\mathrm{terms}\;\mathrm{of}\;\mathrm{degree}\;\delta)^T\,,
\end{equation}
and the polynomial itself can be written in the form
\begin{equation}
  P_j(\x)=\mathbf{p}_j\cdot\mathbf{z},
\end{equation}
where $\mathbf{z}$ contains all of the products of the elements of $\left(\x - \x_0^\prime\right)$ up to the degree $\delta$:
\begin{equation}\label{z_def}
\mathbf{z} = (1,\left(\x_1-\left(\x_0^\prime\right)_1\right),\dots,\left(\x_n-\left(\x_0^\prime\right)_n\right),\dots,\mathrm{terms}\;\mathrm{of}\;\mathrm{degree}\;\delta)\,.
\end{equation}
The values of the parameters $\mathbf{\hat{p}_j}$ that minimize $S_j$ can then be calculated by
\begin{equation}\label{best_params}
  \mathbf{\hat{p}_j}=(Z^TWZ)^{-1}Z^TW\y^\prime_j\,,
\end{equation}
where

\begin{equation}
Z=\left(
\begin{array}{c}
\mathbf{z^\mathrm{NN}_1}\\
\vdots\\
\mathbf{z^\mathrm{NN}_k}
\end{array}
\right),\;\;
W=\left(
\begin{array}{ccc}
w^1& & 0\\
 & \ddots & \\
0 & & w^k
\end{array}
\right),\;\;
\y^\prime_j=\left(
\begin{array}{c}
  \left(\y^\mathrm{NN}_1\right)_j\\
\vdots\\
  \left(\y^\mathrm{NN}_k\right)_j\\
\end{array}
\right)
\end{equation}
and each $\mathbf{z^\mathrm{NN}_i}$ is calculated by substituting $\x^\mathrm{NN}_i$ for $\x$ in (\ref{z_def}).
All of the matrices in Eq.~\eqref{best_params} are independent of the output values of $f$. Consequently, the computationally costly matrix operations, including the inversion, need to be done only once even for a large dimensionality $m$ of the output space of $f$.

\section{Implementation in \textsc{Cosmo++}}\label{app_impl}

The tools described in this paper have been implemented in C++ and made publicly available as a part of the \textsc{Cosmo++} package. In this appendix we describe the different modules available and how to use them. \textsc{Cosmo++} includes full documentation for all of these modules.

\subsection{\kd tree}\label{kdtree_app}

The class \func{KDTree} in \file{kd\_tree.hpp} implements a simple \kd tree functionality. The tree can be built by passing a set of points to the constructor. The additional functionality includes inserting new points using the function \func{insert}, resetting the tree using \func{reset}, re-balancing the tree with the function \func{reBalance}, and finding the $k$ nearest neighbors of a given point using \func{findNearestNeighbors}.
This functionality is enough for using the \kd tree in the emulation algorithm described in this paper. The current implementation uses the Euclidean distance, but it can be extended to include other metrics.

\subsection{Emulator}\label{emulator_app}

The emulator algorithm described in Section \ref{algorithm_sec} has been implemented in the class \func{FastApproximator} in the include file \file{fast\_approximator.hpp}. The constructor takes as input the dimensionality of the input and output spaces of the function $f$ that is being approximated, the training set as two arrays of input and output values of $f$, and the number $k$ of nearest neighbors to use in the approximation.\footnote{The constructor is the equivalent of the function \textsc{InitializeEmul} in the schematic Alg.~\ref{alg:emulator}.} The approximation for a new input point can be performed with the \func{approximate} function, which takes as an input the new point, and the interpolation method (currently linear and quadratic interpolations are implemented), and returns as an output the approximated value of the function.\footnote{The \func{approximate} function is equivalent to \textsc{Emulate} in Alg.~\ref{alg:emulator}.} If desired, the function can also return the $k$ nearest neighbors, distances to the nearest neighbors, and the indices of the nearest neighbors in the training set.

The function \func{approximate} is implemented in two steps, and each of the steps can be called separately as a function. The first step involves finding the $k$ nearest neighbors, which is implemented in the function \func{findNearestNeighbors}. This function takes the new input point and finds the nearest neighbors. If desired, the nearest neighbors themselves, the distances to them, and their indices in the training set can also be returned. After this step, the function \func{getApproximation} can be called, which takes as an input the interpolation method and returns the approximated value. Note that this function can only be called after \func{findNearestNeighbors}. The separation of the approximation algorithm into two steps can be useful in many cases. For example, the error evaluation model, such as in Sect.~\ref{error_sec}, may only depend on the characteristics of the nearest neighbors. So the nearest neighbors can be found as a first step, the error can then be evaluated and only if the error is small enough for the approximation to be acceptable can the interpolation be performed. This approach can save computational time by not calculating the interpolation, which involves matrix operations and can be somewhat slow, unless it is necessary. Another useful application could be obtaining both linear and quadratic interpolations without repeating the nearest neighbor search. If none of this is necessary then the two step approximation can be replaced by a single call to the function \func{approximate}.

New points can be added to the training set with the function \func{addPoint}.\footnote{Equivalent to \textsc{AddPointToTrainSet} in Alg.~\ref{alg:emulator}.} This function adds the new point to the \kd tree, without recalculating the covariance matrix, and checks if the \kd tree is very unbalanced after the insertion, in which case the tree is rebalanced by simply rebuilding it. The criterion for the tree being unbalanced is that the depth becomes $4$ times larger than the depth of a balanced tree with the same number of elements (see Sect.~\ref{algorithm_sec}).

The entire training set can be updated using the \func{reset} function, which takes as an input the new training set. The user can choose whether or not to update the covariance matrix of the input points. For example, if the new training set is a superset of the old one with not many additions, it may not be necessary to update the covariance matrix. Also, updating the covariance matrix will alter the distances to the already existing points, which may further impact the error evaluation model described below. So if one is simply adding a few new points to the training set without recalibrating the error model, then the covariance matrix should not be changed.

\subsection{Error Model}\label{error_model_impl_app}

The simple error model described in Sect.~\ref{error_sec} has been implemented in the class \func{FastApproximatorError} in \file{fast\_approximator\_error.hpp}. The input to the constructor consists of a reference to a \func{FastApproximator} object, a set of test points (the cross-validation points), a function $f_\mathrm{err}$ for which the error is modelled, the \emph{error modelling method}, the target accuracy, and the \emph{decision method}.\footnote{The constructor is equivalent to the \textsc{BuildErrorModel} function in the schematic Alg.~\ref{alg:error}.} We describe each of these parameters in detail below.

The function $\eta:\mathbb{R}^m\rightarrow\mathbb{R}$ takes the output of $f$ and maps it to a single number. This function is used for modelling the error (see Sect.~\ref{error_sec}). For the \emph{Planck} results described in Sect.~\ref{cmb_sec} the fast approximation algorithm outputs a set of $C_l^{TT}$ values as well as a few of the \emph{Planck} likelihoods, which can then be translated to the total \emph{Planck} likelihood. In this case, the decision of whether or not to use the approximation should be based on the error of the total likelihood, rather than the errors of individual $C_l$ values or individual likelihood components.
The function $\eta$ should then take the complex output of the approximation algorithm and translate it to the total likelihood.

As described in Sect.~\ref{error_sec}, the probability distribution $p$ of the error $e_\eta$ is modelled via a distribution that depends on the ratio of $e_\eta$ to a single parameter $\rho$ as in Eq.~\eqref{error_model_example}. The assumption is that the error increases linearly with $\rho$, where $\rho$ can be calculated rapidly for any point in question and its nearest neighbors. Alg.~\ref{alg:error} uses the average Euclidean distance to the nearest neighbors for $\rho$. The \emph{error modelling method} for \func{FastApproximatorError} simply refers to the choice of $\rho$ in general. The following options are currently implemented:
\begin{itemize}
  \item \func{MIN\_DISTANCE} -- The distance to the nearest neighbor.
  \item \func{AVG\_DISTANCE} -- The average distance to the $k$ nearest neighbors.
  \item \func{AVG\_INV\_DISTANCE} -- The weighted average distance to the $k$ nearest neighbors, with weights inversely proportional to the distances, which is useful when performing a weighted interpolation between the nearest neighbors.
  \item \func{SUM\_DISTANCE} -- The geometrical sum of the distances to the nearest neighbors. This is a useful measure of both how far the nearest neighbors are on average and how isotropically they are distributed around the given point.
  \item \func{LIN\_QUAD\_DIFF} -- The difference between linear and quadratic interpolations between the nearest neighbors.
\end{itemize}

The test set (also referred to as the cross-validation set $\Xcv$) is used to estimate the distribution of this ratio. For each point in the test set, the measure $\rho$ is calculated, the approximate and exact values of $\eta$ are calculated, and the ratio $|e_\eta|/\rho$ is obtained. A histogram is then built using these ratios for all of the test points, which is further smoothed with a Gaussian kernel and normalized.

For any new point the parameter $\rho$ is calculated, which is used to rescale the histogram above and obtain the distribution $p$ for the error $e_\eta$ at that point. The decision is made whether or not the approximation is acceptable, by translating the distribution to a single real number and comparing to the target threshold. The choice of the single number representing the distribution is made based on the \emph{decision method} parameter passed to the constructor. The following choices are currently implemented:
\begin{itemize}
  \item \func{ONE\_SIGMA} -- The $68.3\%$ upper bound of the absolute value of the error.
  \item \func{TWO\_SIGMA} -- The $95.5\%$ upper bound of the absolute value of the error.
  \item \func{SQRT\_VAR} -- The square root of the variance of the error.
\end{itemize}

The approximation for a new point is performed using the function \func{approximate}. The approximation itself is performed by passing the point to the underlying \func{FastApproximator}. The return value of this function indicates whether or not the approximation is acceptable. The function can also return the $68.3\%$, the $95.5\%$ upper bounds of the error, as well as the variance and the mean, if any of those are requested.

The error model can be recalibrated with a new test set using the \func{reset} function. The \func{setPrecision} function can be used to change the target accuracy of the model (\emph{i.e.,} the threshold for the error). The distribution for the ratio $|e_\eta|/\rho$ can be obtained using the \func{getDistrib} function.

\subsection{Learn-As-You-Go}\label{layg_app}

The \func{LearnAsYouGo} class in \file{learn\_as\_you\_go.hpp} implements the ``learn-as-you-go'' functionality described in Sect.~\ref{learn_as_you_go_sec}, using the \func{FastApproximator} and \func{FastApproximatorError} classes described above. The purpose of \func{LearnAsYouGo} is to decide for each input point whether or not it should approximate or do the calculation exactly.  If the exact calculation is performed then the result is also added to the training set.

The input parameters to the constructor are the dimensionalities of the input and output spaces of the function $f$ being calculated, a reference to $f$, the error model function $\eta$, the minimum number of points in the training set for which approximations are allowed, the error threshold $\epsilon$, as well as a file name in which the training set can be saved. The calculation is performed using the \func{evaluate} function,\footnote{Equivalent to \textsc{CalculateLAYG} in Alg.~\ref{alg:learnasyougo}.} which will return either an acceptable approximation for the value of $f$ at the given point, or its exact value. If the exact value is calculated and returned then it is automatically added to the training set. The target threshold $\epsilon$ can be changed at any time with the \func{setPrecision} function. The training set can be saved into a file using \func{writeIntoFile} and retrieved using \func{readFromFile}. The progress can be continuously logged into a file by setting the log file using the \func{logIntoFile} function. The log will include the total number of function calls, the number of calls for which the input point already exists in the training set, the number of calls with successful approximations, and the number of calls with unsuccessful approximations, \emph{i.e.}, exact calculations.

An important characteristic of \func{LearnAsYouGo} is that it has a \emph{parallel} implementation. If an application using \func{LearnAsYouGo} is run with many MPI processes, then each process will have its own ``learn-as-you-go'' approximation class. However, all of the processes will periodically share their training sets with each other, so effectively each process will use as a training set all of the exact function calculations by \emph{all of the processes}. The user does not need to do anything special to take advantage of this functionality. All that is required is to run many parallel MPI processes.

\subsection{Accelerated \emph{Planck} Likelihood}

Finally, we use the \func{LearnAsYouGo} class to implement an accelerated ``learn-as-you-go'' \emph{Planck} likelihood in the \func{PlanckLikeFast} class in \file{planck\_like\_fast.hpp}. The interface is similar to the \emph{Planck} likelihood interface in \textsc{Cosmo++}~\cite{Aslanyan:2013opa}. The implementation uses the \func{LearnAsYouGo} class with the $C_l^{TT}$ values and the slower \emph{Planck} likelihoods as the function to be approximated (see Sect.~\ref{cmb_sec}), and the total \emph{Planck} likelihood function to model the error. All of the functionality of \func{LearnAsYouGo} described above, including the parallel implementation, is present in \func{PlanckLikeFast}.
Examples of using \func{PlanckLikeFast} in parameter space samplers are implemented in the \file{test\_mcmc\_planck\_fast.cpp} and \file{test\_multinest\_planck\_fast.cpp} files for Metropolis-Hastings and \textsc{MultiNest} samplers, respectively. Those implementations are used to obtain the results in Sect.~\ref{cmb_sec}.

The constructor takes as an input a pointer to a \func{CosmologicalParams} object, which is used for setting the cosmological parameters, boolean (logical) values defining which of the likelihoods need to be included and whether or not tensor modes are included, as well as the number of points per decade in $k$ space for which the primordial power spectrum is evaluated. Following these parameters, two parameters for the ``learn-as-you-go'' algorithm must be specified. These parameters are the target accuracy (the threshold for the error used to decide whether or not the approximation is acceptable), and the minimum size of the training set before approximations are performed.

The calculation is done through the \func{calculate} function. This function enables the use of the \func{PlanckLikeFast} class in parameter space samplers. Since parameter space sampling is the main application for the ``learn-as-you-go'' likelihood, the detailed interface for calculating each component of the likelihood separately is not implemented here. The regular \emph{Planck} likelihood interface \func{PlanckLikelihood} can be used for that purpose as in Ref.~\cite{Aslanyan:2013opa}.
With this interface, it is very easy to replace the usual \emph{Planck} likelihood code by the accelerated one in any parameter space sampling code. Simply replacing the \func{PlanckLikelihood} object with a \func{PlanckLikeFast} object initialized with all of the necessary parameters will suffice to use the accelerated likelihood code in any parameter space sampler present in \textsc{Cosmo++}.

Since the function being approximated in this case is a likelihood function, we can apply the techniques developed in Sect.~\ref{posterior_error_sec} to estimate the errors of the posterior distributions for the parameters of interest. Our implementation of \func{PlanckLikeFast} allows easy calculation of the posteriors \emph{with errors} as follows. Before starting the parameter space sampling an error log file must be set using the \func{logError} function of \func{PlanckLikeFast}. Once this is set every call of \func{calculate} will output the call information into the error log file, which includes the parameters for the function call, the resulting likelihood, and estimated characteristics of the error, which can then be used to estimate the errors of the posteriors. Whenever the \func{calculate} function does the exact calculation the errors will be zero. There is no modification to the chain files produced by the parameter space samplers. After the sampling is finished, the \func{MarkovChain} class can be used to analyze the chains and obtain the posteriors. The new version of \func{MarkovChain} allows for calculation of errors of the posteriors by simply passing the error log file produced by \func{PlanckLikeFast} as an extra argument to the constructor. In summary, all that is needed for producing posteriors with errors is generating an error log file in \func{PlanckLikeFast} and then passing it to \func{MarkovChain} for analyzing.

\bibliographystyle{JHEP}
\bibliography{citations}

\end{document}